\documentclass[traditabstract]{aa} % for the abstract without structuration 
\usepackage{graphicx}
\usepackage{txfonts}
\usepackage{color}
\usepackage{amstext}

\begin{document}

   \title{Are the gyro-ages of field stars underestimated?}

   \author{G\'eza Kov\'acs\inst{1}
          }

   \institute{Konkoly Observatory, Budapest, Hungary \\
              \email{kovacs@konkoly.hu}
             }

   \date{Received February 18, 2015; Accepted June 16, 2015}

% \abstract{}{}{}{}{} 
% 5 {} token are mandatory
 
  \abstract
%{...}
%{...}
%{...}
%{...}
{By using the current photometric rotational data on eight galactic open 
clusters, we show that the evolutionary stellar model (isochrone) ages 
of these clusters are tightly correlated with the period shifts applied 
to the $(B-V)_0$\textbf{--}$P_{\rm rot}$ ridges that optimally align these 
ridges to the one defined by Praesepe and the Hyades. On the other hand, 
when the traditional Skumanich-type multiplicative transformation is used, 
the ridges become far less aligned due to the age-dependent slope change 
introduced by the period multiplication. Therefore, we employ our simple 
additive gyro-age calibration on various datasets of Galactic field stars 
to test its applicability. We show that, in the overall sense, the gyro-ages 
are systematically greater than the isochrone ages. The difference could 
exceed several giga years, depending on the stellar parameters. Although 
the age overlap between the open clusters used in the calibration and the 
field star samples is only partial, the systematic difference indicates 
the limitation of the currently available gyro-age methods and suggests 
that the rotation of field stars slows down with a considerably lower speed 
than we would expect from the simple extrapolation of the stellar rotation 
rates in open clusters.} 


\titlerunning{Are the gyro-ages of field stars underestimated?}
\authorrunning{G\'eza Kov\'acs}
   \maketitle
%
%________________________________________________________________

%%%%%%%%%%%%%%%%%%%%%%%
% SECTION 1
%%%%%%%%%%%%%%%%%%%%%%%
%
\section{Introduction} 
Gyrochronology (the determination of stellar ages from their rotation periods 
and colors) has gained considerable popularity in recent years, largely 
due to the speedily accumulating observational data on open clusters. These 
data suggest that stars, after several ten million years of formation, 
settle on a fairly well-defined ridge in the color$-$rotation period 
($P_{\rm rot}$) diagram. By comparing cluster data of various ages, 
it turned out that the height of these ridges (i.e., the rotation periods 
at each color) increases as the cluster is aging. This property has 
first been recognized by Skumanich~(\cite{skumanich1972}) (see also 
Kraft~\cite{kraft1967}) and has later been elaborated by many authors both 
observationally (e.g., Barnes~\cite{barnes2003}) and theoretically (e.g., 
Kawaler~\cite{kawaler1988}). Although differing in details, it is widely 
accepted that the slowing-down of the rotation is due to the angular 
momentum loss by magnetized stellar wind (as first described by 
Schatzman~\cite{schatzman1962}), and the rotation period is scaled as 
the square root of the stellar age (as suggested first by Skumanich). 
A comprehensive description of the current status of the field of stellar 
rotation can be found, for example, in Bouvier~(\cite{bouvier2013}). 

The success of the applicability of gyrochronology depends on various 
factors, most importantly on the validity of the relation derived from 
open clusters for other stars. Based on a rather limited sample, 
Barnes~(\cite{barnes2009}) have shown that both the chromospheric and 
the isochrone ages are considerably greater than the gyro-ages derived 
from his formulae. Similarly, Brown~(\cite{brown2014}) investigated a 
more extended sample of transiting extrasolar planet host stars and 
found hints of this effect. Two very recent papers seem to further 
strengthen this observation. Angus et al.~(\cite{angus2015}) performed 
a Monte Carlo Markov chain analysis by using $310$ asteroseismic 
targets from the archive of the Kepler satellite, a few well-studied 
fields stars from other earlier works, and data on two open clusters. 
They calibrated the formula of Barnes~(\cite{barnes2003}) on this 
merged dataset. They found that the marginalized likelihoods of the 
gyro-parameters exhibit multiple maxima. The authors suggest the 
presence of multiple rotation-color-age populations and raise concerns 
for the currently applied method of gyrochronology. In a further paper, 
Maxted, Serenelli, \& Southworth~(\cite{maxted2015}) found strong 
evidence for the younger gyro-ages of many of the $28$ extrasolar 
planet host stars in their sample (largely discovered by ground-based 
surveys). They searched for a possible cause of the discrepancy within 
the framework of tidal interaction between the planet and the host, 
but they found no compelling evidence for a relation between the 
gyro-age and the computed timescale of tidal interaction. 

There are also technical details, including the transformation of the 
color and period values to stellar ages. This is usually done by the 
type of formulae introduced by Barnes~(\cite{barnes2003}), where, by 
maintaining the Skumanich-type age dependence, the color dependence 
of the period is represented by a multiplicative factor that entirely 
depends on the color. We show that this representation is 
suboptimal because it does not lead to the cleanest average 
color$-P_{\rm rot}$ ridge when, using the prescribed time-dependence, 
all periods are transformed to the same age. Yet another, equally 
important question is if the target star has already reached (and is 
still in) the rotationally settled state (corresponding to sequence 
`I' in the nomenclature of Barnes~\cite{barnes2003}). 
Open clusters are vital objects to define this state since their members 
are assumed to be coeval, leading to a topographically well-defined 
ridge structure in the color$-P_{\rm rot}$ plane. However, for 
individual targets we do not have any criterion to decide whether they  
are in the rotationally settled state, except that we assume that for 
their estimated ages they are. This is a general problem
in the applicability of gyrochronology.    
 
The purpose of this paper is to derive an updated relation between the 
color, rotation period, and age of rotationally settled stars in 
open clusters and employ this relation on various independent datasets 
 to test the new formula against isochrone ages and, if possible, 
improve the gyro-method and extend its applicability. Except for
using the isochrone ages, we stay within a strictly empirical 
framework throughout the paper.

%%%%%%%%%%%%%%%%%%%%%%%
% SECTION 2
%%%%%%%%%%%%%%%%%%%%%%%
%
\section{New color-period-age formula}
The `I' branches of the cluster rotational data are commonly fitted by the 
following expression (Barnes~\cite{barnes2003})

%################
%  Eq. (1)
%################
%
\begin{equation}
P = g(t)f(B-V) \hskip 2mm , 
\end{equation}
where the age-dependent factor $g(t)$ is approximated by a power-law 
expression of $t^n$ with $n\approx 0.5$ in a broad agreement with the 
Skumanich-law. The color-dependent factor is also represented as a 
power law: $a(B-V-c)^b$, where $a$, $b,$ and $c$ are constant parameters 
determined by some properly chosen cluster; $B-V$ is assumed to be 
free of reddening. 

\begin{figure}
 \vspace{0pt}
 \includegraphics[angle=-90,width=85mm]{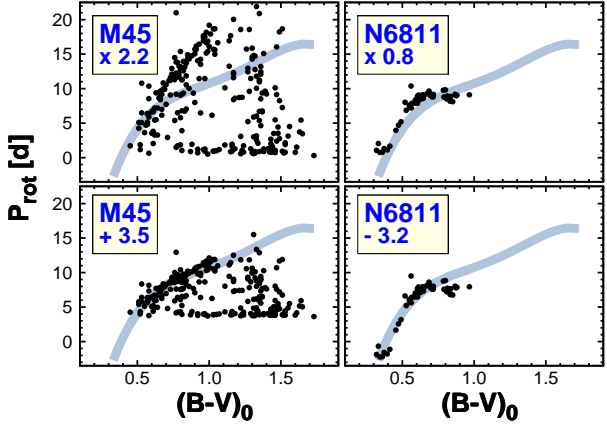}
 \caption{Multiplicative vs. additive ridge alignments. Upper panels 
          show the result of the exact Skumanich-type period scaling with 
	  $P_{\rm rot}\rightarrow P_{\rm rot}\sqrt{t_{\rm M44}/t_{\rm cluster}}$, 
          where $t$ stands for cluster ages. The scaling factors are 
          displayed in the boxes. The lower panels show the same test 
          with the additive scaling of 
          $P_{\rm rot}\rightarrow P_{\rm rot}+c$, where $c$ is the 
          period shift as shown in the corresponding boxes. The fiducial 
          ridge (see Eq.~(3)) is plotted as a light color continuous line.}
\label{mult-vs-add}
\end{figure}

In a brief test of this multiplicative age dependence, we transformed 
two clusters (Pleiades (M45) and NGC~6811) to the fiducial ridge line 
(determined by Praesepe (M44) and the Hyades -- see Sect.~2.2). 
The multiplicative and additive period transformations are shown
in the upper and lower panels of Fig.~\ref{mult-vs-add}. The change in 
slope for the younger cluster M45 in the case of the multiplicative 
transformation is clear (see also Cargile et al.~\cite{cargile2014}, 
their Fig. 12). For NGC~6811 there is not much difference between the 
two types of transformation. However, we see (although the color range 
is rather short) that in both cases there is a slight downward trend 
with respect of the fiducial ridge. It might be that the steeper slope 
for young clusters changes toward a milder slope and eventually reverses 
for older clusters. Unfortunately, NGC~6811 is the only available cluster 
with high-quality rotational periods at $\sim 1$~Gyr, so the empirical 
confirmation of the evolution of slope needs to await future observations 
of older clusters. Because of the considerably better performance of 
the additive period transformation in the younger age range, in the next 
sections we therefore derive a new calibration of the gyro-age relation 
based on the additive method.  

%################
% Figure 1
%################
%

%%%%%%%%%%%%%%%%%%%%%%%
% Section 2.1
%%%%%%%%%%%%%%%%%%%%%%%
%
\subsection{Calibrating datasets}
We chose eight recently observed clusters with reliable colors and 
periods. The basic properties of these clusters that are relevant for 
this paper are listed in Table~\ref{clusters}. The ages span the range 
between $\sim0.1$ and $\sim1$~Gyr. Unfortunately, for older clusters 
(more relevant for the field stars) we do not yet have good rotation 
period data. For example, for one of the oldest open clusters, 
M67, the available rotation data are too sparse to be considered useful 
in the present context (e.g., Canto Martins et al.~\cite{martins2011}; 
Stassun et al.~\cite{stassun2002}). For NGC~6819 
(age $\sim 2.5$~Gyr, see Balona et al.~\cite{balona2013b}), it is hard 
to reconcile the color-period plot assuming that the cluster members are 
coeval. For NGC~6866 (age $\sim 0.65$~Gyr, see Balona et al.~\cite{balona2013a}), 
the same diagram is cleaner but still confusing (nearly constant period 
of $\sim 10$~days from $(B-V)_0\approx 0.5$ with a nearly uniform 
downward scatter between $\sim 2$ and $\sim 10$~days).     

Although from both an observational and theoretical point of view, the 
use of $(B-V)_0$ as the color coordinate in the rotational studies is 
not necessary the best one, it is commonly used. Therefore, we follow  
this practice here as well and use the previously published $(B-V)$ 
colors whenever possible as given in the corresponding papers where 
the rotational periods were published. Four clusters (M34, M35, M45, and 
Blanco~1) fall into this category. For M44, we also use the $(B-V)$ colors 
published in our period source (i.e., in Kov\'acs et al.~\cite{kovacs2014}) 
but we note that the colors originate from the APASS database (via the UCAC4 
catalog, see Zacharias et al.~\cite{zacharias2013}). For M37 we use the 
colors given in the period source (Hartman et al.~\cite{hartman2009}), 
which are based on the cluster study of Kalirai et al.~(\cite{kalirai2001}), 
however. For NGC~6811 we cross-correlate the list of rotational variables 
of Meibom et al.~(\cite{meibom2011a}) with the photometric table of 
Janes et al.~(\cite{janes2013}). This yields $58$ objects from the original 
$71$ objects of Meibom et al.~(\cite{meibom2011a}). For the Hyades we use 
the compilation of colors as given in the period source by 
Delorme et al.~(\cite{delorme2011}). Six stars in this source do not have 
color values. We checked the APASS database for these objects and found 
that five of these have fine $B$, $V$ measurements. Before entering these 
values, we cross-correlated the other stars of 
Delorme et al.~(\cite{delorme2011}) with the APASS database and found that 
on the average the APASS $(B-V)$ indices are $0.04$~mag bluer than the 
ones given in Delorme et al.~(\cite{delorme2011}), that is, 
$(B-V)_{\rm compilation}=(B-V)_{\rm APASS}+0.04$. By applying this color 
shift to the APASS colors of the five variables mentioned above, we finally 
compiled a sample with $61$ variables for the Hyades. 

%================
% Table 1
%================
%
\begin{table}
 \centering
 \begin{minipage}{140mm}
  \caption{Calibrating clusters}
  \label{clusters}\scalebox{0.85}{
  \begin{tabular}{lccrl}
  \hline
   Cluster  & E(B-V) & Age [Gyr] & $N_{\rm LC}$ & Source\\
 \hline
 Blanco~1 &  $0.02$  & $0.117\pm0.017$ &  33 & C09, C10, C14 \\
 M45      &  $0.04$  & $0.135\pm0.015$ & 251 & B14, B14, H10 \\
 M35      &  $0.20$  & $0.180\pm0.020$ & 418 & K03, K03, M09 \\
 M34      &  $0.07$  & $0.220\pm0.030$ & 118 & C79, M11b, M11b \\
 M37      &  $0.23$  & $0.550\pm0.030$ & 575 & H08, H08, H09 \\
 Hyades   &  $0.00$  & $0.625\pm0.050$ &  61 & T80, P89, D11 \\
 M44      &  $0.03$  & $0.665\pm0.011$ & 180 & T06, B14, K14 \\
 NGC6811  &  $0.07$  & $1.000\pm0.170$ &  58 & J13, J13, M11a \\
\hline
\end{tabular}}
\end{minipage}
\begin{flushleft}
\underline{Notes:} 
All ages result from isochrone fits. $N_{\rm LC}=$ number of light 
curves; source: papers used for reddening, age and rotation periods.  
We assigned arbitrary age errors to M34 and M35. For these two clusters 
we used the published dereddened colors. For M35 
Meibom, Mathieu \& Stassun~(\cite{meibom2009}) did not specify the 
reddening correction used. See text on the sources of the colors used 
for the clusters entered in this table. \\   
{\em $E(B-V)$ source:} C09=Cargile, James \& Platais~(\cite{cargile2009}),
                       B14=Bell et al.~(\cite{bell2014}), 
                       K03=Kalirai et al.~(\cite{kalirai2003}),
                       C79=Canterna, Crawford \& Perry~(\cite{canterna1979}),
                       H08=Hartman et al.~(\cite{hartman2008}),
                       T80=Taylor~(\cite{taylor1980}),
                       T06=Taylor~(\cite{taylor2006}),
                       J13=Janes et al.~(\cite{janes2013}) \\
{\em Age source:}      C10=Cargile, James \& Jeffries~(\cite{cargile2010}) (from their Fig.~3), 
                       B14=Bell et al.~(\cite{bell2014}),
                       K03=Kalirai et al.~(\cite{kalirai2003}),
                       M11b=Meibom et al.~(\cite{meibom2011b}),
                       H08=Hartman et al.~(\cite{hartman2008}),
                       P98=Perryman et al.~(\cite{perryman1998}), 
                       J13=Janes et al.~(\cite{janes2013}) \\
{\em Period source:}   C14=Cargile et al.~(\cite{cargile2014}),
                       H10=Hartman et al.~(\cite{hartman2010}), 
                       M09=Meibom et al.~(\cite{meibom2009}),
                       M11b=Meibom et al.~(\cite{meibom2011b}),
                       H09=Hartman et al.~(\cite{hartman2009}),
                       D11=Delorme et al.~(\cite{delorme2011}),
                       K14=Kov\'acs et al.~(\cite{kovacs2014}),
                       M11a=Meibom et al.~(\cite{meibom2011a})
\end{flushleft}
\end{table}

%%%%%%%%%%%%%%%%%%%%%%%
% Section 2.2
%%%%%%%%%%%%%%%%%%%%%%%
%
\subsection{Fiducial color--period ridge, age scaling}
As we have discussed in the introduction of this section, the strong 
variation in the steepness of the color-period ridges when a 
Skumanich-type multiplicative period transformation is used suggests 
that there is a need for some other type of transformation if we assume 
that these ridges are related through some simple, few-parameter 
transformation. Visual inspection of the various cluster data suggests 
that a simple vertical (i.e., period) shift may substantially 
improve the ridge alignment. Therefore, we assumed that there 
exists a {\em fiducial ridge} that yields a better representation of 
the data simply by an optimum shift of this ridge, that is, the rotation 
periods of the stars on the ridge of each cluster can be represented 
by the following formula

%################
%  Eq. (2)
%################
%
\begin{equation}
P_{\rm rot} {\rm (age, color)} =  P_{\rm rot}^{\rm fiducial} {\rm (color)} - c{\rm (age)} \hskip 2mm ,\end{equation}
where we assumed that the additive constant is primarily a function of 
age. The derivation of the additive gyro-age relation requires determining 
the fiducial ridge and the cluster-by-cluster period shifts. In the earlier 
version of the paper we employed an iterative scheme of least squares with 
data point density as weights to derive the main fiducial ridge as the best 
polynomial approximation for the high-density `I' branch part of the 
$(B-V)_0-P_{\rm rot}$ diagram. The derived fiducial ridge was rather close 
to the one spanned by Praesepe. In part due the stimulation of the referee 
report, we therefore decided to fix the fiducial ridge as a polynomial fit 
to the merged data of Praesepe and the Hyades.\footnote{We added the Hyades 
because the two clusters are very similar in all aspects, including 
their ages, therefore, except for proper correction for - their otherwise 
low - reddening, merging does not require any additional special treatment 
of the data.} Some details of the fitting procedure are given in Appendix~A. 
Here we only give the finally accepted fourth-order polynomial expression 

%################
%  Eq. (3)
%################
%
\begin{eqnarray}
P_{\rm rot}^{\rm fiducial}   
= & - &35.51 + 153.07(B-V)_0 - 201.13(B-V)_0^2 \nonumber \\
& \pm & \phantom{3}7.71 \pm \phantom{1}33.07 \phantom{(B-V)_0} \pm \phantom{2} 50.84 \nonumber \\
& + & 120.60(B-V)_0^3 - 26.28(B-V)_0^4 \nonumber \\
& \pm & \phantom{1}33.35 \phantom{(B-V)_0} \pm \phantom{2} 7.91 \hskip 2mm .
\label{fiducial_pol}
\end{eqnarray}
This polynomial fits the periods of the two clusters with 
$\sigma_{\rm fit} = 1.088$~days. The errors listed are $1\sigma$ formal 
errors.  

\begin{figure}
 \vspace{0pt}
 \includegraphics[angle=-90,width=85mm]{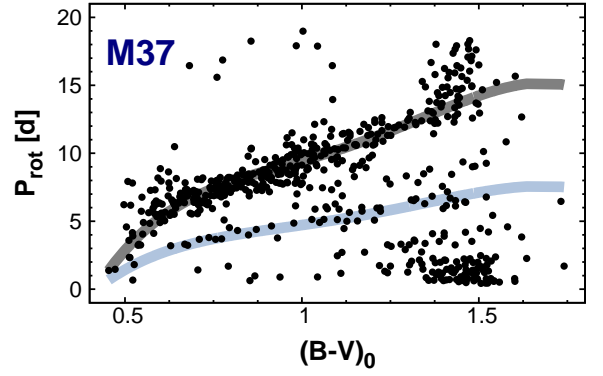}
 \caption{Example of the period adjustment based on the harmonic 
          ambiguity of the estimated rotation periods from the 
          frequency spectra of the light curves 
          (Hartman et al.~\cite{hartman2009}). The darker gray line 
          shows the fiducial ridge line shifted to the cluster `I' sequence.  
          The lighter gray line was obtained by halving the periods 
          corresponding to this cluster ridge line. The uncorrected 
          (i.e., originally published) periods are displayed as 
          black dots.}
\label{sub-harmonic}
\end{figure}

We need to note here that the clusters used to derive this fiducial 
ridge exhibit fairly clean color$-P_{\rm rot}$ diagrams, including 
the lack of major ambiguities concerning the relation of the true rotational 
period to the one determined by the highest peak in the frequency spectrum. 
There is a generic degeneracy in this respect in all photometrically 
determined rotation periods. They are ambiguous toward integer multiple 
periods due to possible special spot positions and numbers and also for 
aliasing in the case of ground-based observations. We show a clear example 
of the period ambiguity in Fig.~\ref{sub-harmonic}. The number of ambiguous 
periods changes from cluster to cluster and have some influence on the 
resulting period shifts. However, our experience shows that although this 
correction is important in principle, in practice (in part due to the 
relatively small number of ambiguous cases) it does not have a significant 
effect on the robust period shift estimation described below. 

%################
% Figure 2
%################
%

Several ways are possible to find the fiducial ridge line that best fits an ensemble of points that 
contains the cluster ridge line (sequence `I') as a subset. Manual selection of outliers is one possibility, 
weighting with the density of the data points is another. Here we resorted 
to a robust fitting method that is based on a special kernel function employed 
in the least-squares fit. The kernel will automatically put less weight 
on outliers, and we do not need to decide on a case-by-case basis whether the 
given data point is an outlier or not. 

From the several kernels available in the literature, we chose the one 
introduced by German \& McClure~(\cite{german1987}). The GM kernel is widely 
used in various robust fit problems, including pattern recognition 
(e.g., Yang et al.~\cite{yang2014}). Accordingly, we minimized the following 
expression to find the best-fitting fiducial ridge

%################
%  Eq. (4)
%################
%
\begin{eqnarray}
E               & = & \sum_{i=1}^n G(\delta P(i)) \nonumber \\ 
G(\delta P(i))  & = & (\delta P(i))^2 \over a+(\delta P(i))^2 \nonumber \\
\delta P(i)     & = & P_{\rm rot}^{\rm fiducial}(i) - P_{\rm rot}(i) - c \hskip 2mm .
\label{gm-ls}
\end{eqnarray}This least-squares condition with the GM kernel is 
equivalent to a weighted ordinary least-squares condition with Cauchy 
weights. We used simple scanning to find the best-fitting period shift 
parameter $c$. The error of $c$ is estimated as 
$\sigma^2(c)=\sum_{i=1}^n w(i)(\delta P(i))^2/(n\sum_{i=1}^n w(i))$,  
where $w(i)=1/(a+(\delta P(i))^2)$. 
The method has a single free parameter $a$ that can be tuned to be more 
($a$ is small) or less ($a$ is large) sensitive to outliers. We found that 
$a=1$ yields a perfect performance with accurate and robust identification 
of the `I' sequences in each cluster, with {\em all data} listed in 
Table~\ref{clusters} left in the datasets for each cluster. An example 
of the cluster fit is shown in Appendix A. The optimum period shifts with 
their errors are summarized in Table~\ref{cluster-shifts}. 

\begin{figure}
 \vspace{0pt}
 \includegraphics[angle=-90,width=90mm]{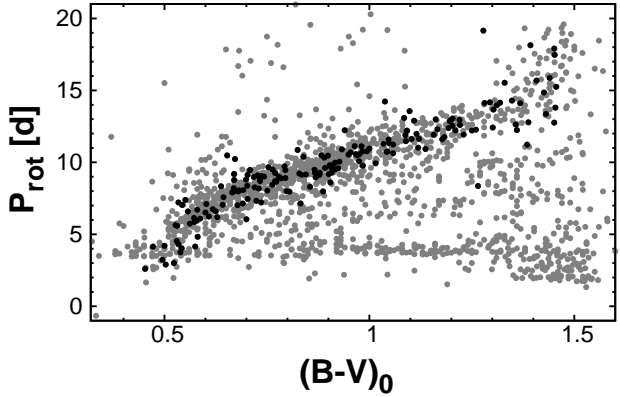}
 \caption{Merged data of the eight clusters of Table~\ref{clusters} 
          after applying the shifts to each cluster as given in 
          Table~\ref{cluster-shifts}. Data from M44 (Praesepe) are 
	  shown as black dots.}
\label{ridge-add}
\end{figure}

\begin{figure}
 \vspace{0pt}
 \includegraphics[angle=-90,width=90mm]{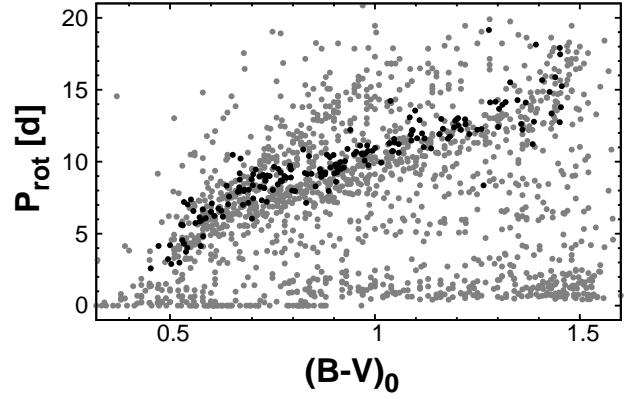}
 \caption{Merged data of the eight clusters after multiplying the 
          cluster periods by the factors given by the exact Skumanich-law 
          (i.e., $P_{\rm tranf}=P_{\rm obs}\sqrt{AGE(\rm M44)/AGE(\rm clus)}$.) 
          The cluster ages are given in Table~\ref{clusters}. Data from 
          M44 (Praesepe) are shown by black dots.}
\label{ridge-mult}
\end{figure}

To illustrate the difference between the additive and multiplicative 
(Skumanich-type) period transformations on the full sample of open 
clusters used in this paper, in Figs.~\ref{ridge-add} and \ref{ridge-mult}
we plot all data points after applying these transformations. 
It is clear that the additive transformation leads to a tighter pattern 
for the `I' sequence and thereby allows a more reliable investigation 
of the age dependence of the rotation periods throughout the $(B-V)_0$ 
color range of $\sim (0.5-1.4)$.  

%================
% Table 2
%================
%
\begin{table}
\setlength{\tabcolsep}{4pt}
 \centering
 \footnotesize
  \begin{minipage}{200mm}
  \caption{Period shifts for the calibrating clusters}
  \label{cluster-shifts}
  \begin{tabular}{rrrrrrrr}
  \hline
   Blanco~1  & M45  & M35  & M34  & M37  & Hyades & M44   & N6811 \\
 \hline\small
   $3.08$      & $3.48$ & $3.37$ & $2.53$ & $1.31$ &  $0.05$  & $0.01$ & $-3.16$ \\
$\pm0.17$    & $\pm0.08$ & $\pm0.07$ & $\pm0.11$ & $\pm0.05$ &  $\pm0.12$  & $\pm0.05$ &  $\pm0.11$ \\
\hline
\end{tabular}
\end{minipage}
\begin{flushleft}
\underline{Notes:}
The shifts (parameter $c$ in Eq.~(2)) are given in days. Equations~(2) 
and (3) can be used to predict the ridge periods for each cluster.   
\end{flushleft}
\end{table}

%################
% Figure 3
%################
%

%################
% Figure 4
%################
%

%################
% Figure 5
%################
%
\begin{figure}
 \vspace{0pt}
 \includegraphics[angle=-90,width=90mm]{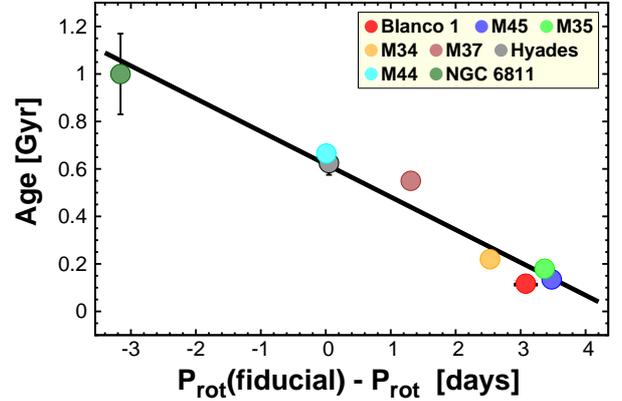}
 \caption{Linear regression of the optimum period shift parameters 
          to the cluster isochrone ages (see Tables~\ref{clusters} and 
          ~\ref{cluster-shifts} and Eq.~(5)). Except for NGC~6811, the 
          estimated errors (both for the isochrone ages and for the 
          period shifts) are smaller or nearly of the size of the dots. 
          The standard deviation of the fit is $0.16$~Gyr.}
\label{pshift-age}
\end{figure}

With the optimum period shifts we are in the position to investigate 
the functional dependence of the shift on cluster age. An inspection 
of this plot (see Fig.~\ref{pshift-age}) by eye clearly shows that a simple 
linear correlation should yield a fairly accurate description of the 
functional dependence (at least at the current stage, with eight data 
points at hand). The linear regression yields the following expression 
and formal error estimate  

%################
%  Eq. (5)
%################
%
\begin{eqnarray} 
          {\rm AGE}_{\rm gyro} & = & a_0 + a_1\Delta P \nonumber\\
\sigma^2({\rm AGE}_{\rm gyro}) & = & K_{11} + K_{22}(\Delta P)^2 + a_1^2\sigma^2(\Delta P) + 2K_{12}\Delta P \nonumber \\
                   \Delta P & = & P_{\rm rot}^{\rm fiducial} - P_{\rm rot}  \nonumber \\
                        a_0 & = & \, \, \, 0.620, \quad\quad a_1=-0.138 \nonumber \\ 
                     K_{11} & = & \, \, \, 0.004682, K_{22}=0.000734 \nonumber \\
                     K_{12} & = & -0.000978
.\end{eqnarray}

\noindent 
The fiducial period $P_{\rm rot}^{\rm fiducial}$ can be easily evaluated 
for any star given its dereddened color and using Eq.~(3). This
means that Eq.~(5) 
is directly applicable to the estimation of the gyrochronological age of 
any star with known color and rotation period. We 
note that the error formula was derived by assuming a uniform isochrone age error as given 
by the standard deviation of the fit.  

We emphasize that this equation was derived by using the {\em isochrone} 
ages of the calibrating clusters. Therefore, we expect this formula to 
yield a fair approximation of the isochrone age of any (rotationally settled) 
target within the calibrating parameter range and hopefully beyond. 

\begin{figure}
 \vskip -0mm
 \includegraphics[angle=-90,width=90mm]{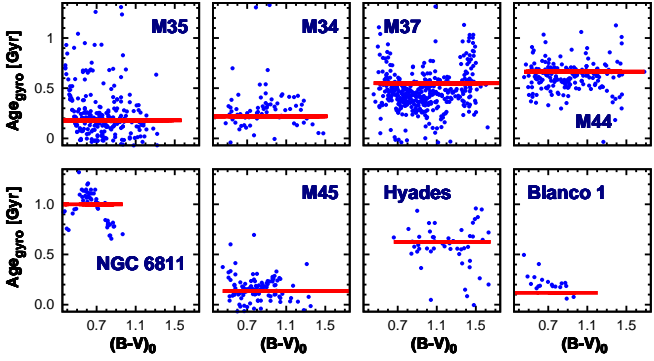}
 \vskip -10mm
 \caption{Predicted individual gyro-ages (dots) by using Eqs.~(3) and (5)  
          and the adopted isochrone cluster ages of Table~\ref{clusters} 
          (lines).}
\label{gyro-iso-clusters}
\end{figure}

It is interesting to compare the derived star-by-star gyro-ages for each 
cluster with their isochrone ages. The result of this comparison is shown 
in Fig.~\ref{gyro-iso-clusters}. Although there are clusters (i.e., M37, 
M45, NGC~6811) with some observable trend in the run of the individual 
gyro-ages, the overall fit is satisfactory. This figure also highlights 
the relatively large errors we may expect when the gyrochronological method 
is employed on individual targets (even if there was a way to ascertain 
that they are in the rotationally settled state). For comparison, in 
Appendix B we show the same plot by using the gyro-age formula of 
Angus et al.~(\cite{angus2015}). Their formula is based on the functional 
form of period and color dependence of Barnes~(\cite{barnes2003}). Because 
of the multiplicative nature of the age-period dependence, we see larger 
systematic variations in the estimated ages than in the case of the 
additive dependence.   

%################
% Figure 6
%################
%

%%%%%%%%%%%%%%%%%%%%%%%
% SECTION 3
%%%%%%%%%%%%%%%%%%%%%%%
%
\section{Tests on independent datasets}
Although the gyro-age formula derived in Sect.~2 is expected to have 
a limited applicability, it is important to see how broad this limit 
is. As mentioned, we cannot investigate whether the stars to be tested are 
in the rotationally settled state. This, by itself, introduces a great 
deal of uncertainty. The isochrone ages are also erroneous, sometimes 
so excessively (e.g., for K and M dwarfs) that individual ages rarely 
have any value. Nevertheless, such a test (if employed on a sufficiently 
large sample) may tell us something about the applicability of the gyro-age 
method and may also shed light on the evolution of the rotation of various 
stellar populations.

%%%%%%%%%%%%%%%%%%%%%%%
% Section 3.1
%%%%%%%%%%%%%%%%%%%%%%%
%
\subsection{Bright field stars and the Sun}
Valenti \& Fisher~(\cite{valenti2005}) published accurate spectroscopic 
parameters (including $v_{\rm rot}\sin i$ values) for over one
thousand nearby 
bright, mostly main-sequence F--K stars. We adopted their isochrone ages 
(based on the Yonsei-Yale stellar evolution models of 
Demarque et al.~\cite{demarque2004}) and used their mass and gravity 
values to compute the stellar radii. From these and the rotation 
velocities, we estimated the rotation periods (in [days]): 
$P_{\rm rot}=50.6R/(v_{\rm rot}\sin i)$, where the stellar radius 
$R$ is in solar units and the rotation velocity $v_{\rm rot}\sin i$ is 
in [kms$^{-1}$]. From the effective temperature, gravity, and metallicity 
given by Valenti \& Fisher~(\cite{valenti2005}), we can estimate 
$(B-V)_0$ with the aid of the formula of Sekiguchi \& Fukugita~(\cite{sekiguchi2000}) 

%################
%  Eq. (6)
%################
%
\begin{eqnarray}
(B-V)_0 & = & -813.3175 + 684.4585\log T_{\rm eff} \nonumber \\
        & - &  189.923(\log T_{\rm eff})^2 + 17.40875(\log T_{\rm eff})^3 \nonumber \\
        & + &    1.2136[Fe/H] + 0.0209[Fe/H]^2 \nonumber \\ 
        & - &     0.294[Fe/H]\log T_{\rm eff} -1.166\log g \nonumber \\ 
        & + &    0.3125\log g\log T_{\rm eff} \hskip 2mm . 
\end{eqnarray}
With these parameters we can evaluate the gyro-age for each 
star and compare it with the corresponding isochrone age. In a similar 
representation we can plot the isochrone age as a function of 
$P_{\rm rot}(\rm fiducial)-P_{\rm rot}$ and compare it with the gyro-age 
estimate of Eq.~(5). In the data preparation we excluded objects with 
rotational velocities lower than $0.1$~kms$^{-1}$ and also those that 
did not pass our criterion of isochrone fitting (see Sect.~3.3). 
Finally, we had a sample with $934$ stars from the original sample of $1039$ 
stars of Valenti \& Fisher~(\cite{valenti2005}). We also note that for 
demonstrative purpose we trimmed the period shift$-$age plotting area to 
the space of interest, thereby excluding some small fraction of objects. 
This does not affect our conclusion in any way because on the one hand, 
these objects have extreme parameter values that are largely irrelevant 
for the effect we investigate, and on the other hand, their observable 
and derived parameters (e.g., age, rotation period) are also generally 
inaccurate.  

\begin{figure}
 \vskip -0mm
 \includegraphics[angle=-90,width=90mm]{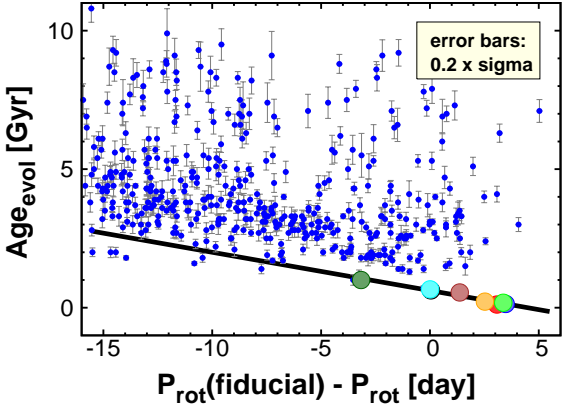}
 \vskip -0mm
 \caption{Period shift vs. evolutionary (i.e., isochrone) ages for 
          the field star sample of Valenti \& Fisher~(\cite{valenti2005}) 
          (blue dots) and for the eight open cluster ridges as 
          presented in this paper (large colored dots). The linear 
          regression to the cluster data is shown by the black line.}
\label{evol-gyro-valenti}
\end{figure}

In Fig.~\ref{evol-gyro-valenti} we plot the isochrone ages given by 
Valenti \& Fisher~(\cite{valenti2005}) as a function of the rotational 
period shift. For reference, we also overplot the corresponding cluster 
values.  Clearly, there is a striking difference between the isochrone 
and gyro-ages. Nearly all isochrone ages are greater than the gyro-ages. 
If we focus only on the more densely populated part of the plot, we see 
that the overall difference is $2$~Gyr. Unfortunately, the region below 
$1$~Gyr (where the calibration was performed) is rather sparsely populated. 
Nevertheless, it seems that even in this regime the isochrone ages carry 
the same property as in the non-calibrated (older age) regime. We note 
that correcting the spectroscopic rotation velocities for the overall 
aspect effect (e.g., Nielsen et al.~\cite{nielsen2013}) exacerbates the 
situation because it leads to higher rotational velocities, shorter 
periods, and therefore to even younger gyro-ages. Furthermore, possible 
systematic bias in the rotation velocities might occur at low-rotation 
rates when the rotational broadening is similar to other broadening effects 
(e.g., stellar macroturbulence). However, these stars do not contribute 
in an important way to the discrepancy above, since more than half of 
the sample has $v\sin i > 2$~km$^{-1}$.   

The case of the Sun is special because we know both its age and its rotation 
period, together with other physical parameters. Taking the equatorial 
rotation period, we derive that the gyro-age of the Sun is $3.74\pm0.65$~Gyr. 
Again, the gyro-age is $0.9$~Gyr short relative to the accurately known 
age of $4.6$~Gyr (which happens to be close to the actually computed 
isochrone age of $4.4$~Gyr, based on the Yonsei-Yale models discussed 
in Sect.~3.3.) 

To show that this age discrepancy is not unique for the additive 
period$-$age scaling introduced in this paper, we applied two different 
period$-$color$-$age calibrations on the above dataset. The result  
(presented in Appendix B) clearly shows that the discrepancy is generic.

%################
% Figure 7
%################
%

%%%%%%%%%%%%%%%%%%%%%%%
% Section 3.2
%%%%%%%%%%%%%%%%%%%%%%%
%
\subsection{Transiting extrasolar planet host stars}
The host stars of extrasolar planets are usually very deeply studied 
objects because it is important to accurately determine the stellar 
parameters to derive the planet parameters. In addition, if the planet 
is also transiting planet, the average stellar density is rather tightly 
constrained by the basic orbital parameters of the planet 
(Seager \& Mall\'en-Ornelas~\cite{seager2003}; 
Sozzetti et al.~\cite{sozzetti2007}). As a result, these stars are very 
useful targets for testing the gyro-age method. 

We compiled the relevant physical parameters of $147$ bright transiting 
planet host stars from the literature. All of these systems have been 
discovered by ground-based surveys. All ages used are stellar model 
(isochrone) ages, and they have been derived mainly from the Yonsei-Yale 
models. Unfortunately, the rotation periods are still based on the 
spectroscopic $v_{\rm rot}\sin i$ values, since there are rather few host 
stars with reliable direct photometric rotation periods. 

\begin{figure}
 \vskip -0mm
 \includegraphics[angle=-90,width=90mm]{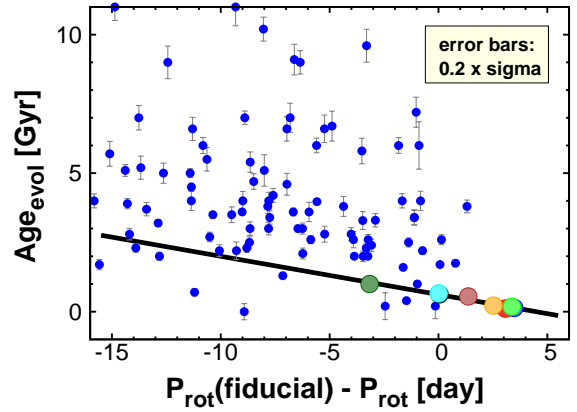}
 \vskip -0mm
 \caption{As in Fig.~\ref{evol-gyro-valenti}, but for the sample of 
          hot-Jupiter host stars, discovered by ground-based surveys.}
\label{evol-gyro-hj}
\end{figure}

The result is plotted in Fig.~\ref{evol-gyro-hj} in the same fashion as 
before. Unfortunately, some stars, otherwise possessing accurately determined 
stellar parameters, had to be omitted because of their extreme parameters. 
For instance, WASP-33 is a rare A-type host star, with a high rotation rate 
of $90$~kms$^{-1}$, yielding a $P_{\rm rot}$ value of $0.79$~days. The 
isochrone age is fairly well constrained with an upper limit of $0.4\pm0.3$~Gyr 
(Collier Cameron et al.~\cite{collier2010}; Kov\'acs et al.~\cite{kovacs2013}). 
Although the fast rotation is consonant with its young age, we cannot verify 
this with our gyro-age formula because the target is outside its validity. 
For similar reasons we excluded five stars. As in the test of the 
Valenti \& Fisher~(\cite{valenti2005}) dataset, we also focus on a limited 
area of the period shift$-$age space (and again, only a small fraction 
of objects were excluded, which does not alter our conclusion). 

Although the data are sparser than for the large nearby star survey of 
Valenti \& Fisher~(\cite{valenti2005}), the {\em same} effect is still  
well visible: the highly significant excess of stars dated older by the 
isochrone age determination. With the overall more accurate isochrone ages 
for these bright, well-studied stars, the discrepancy between the two 
types of age determination is reaffirmed. 

Similarly to the test presented on the Valenti \& Fisher~(\cite{valenti2005}) 
dataset in Sect. 3.1, here we refer to Appendix B, where we compare our 
gyro-ages with those recently derived by Maxted et al.~(\cite{maxted2015}) 
on a sample of limited-number extrasolar host stars with measured rotation 
periods. In spite of our very different approach, the two types of gyro-ages 
correlate very well, supporting their conclusion on the shorter gyro-ages 
for their planet host sample.

%################
% Figure 8
%################
%

%%%%%%%%%%%%%%%%%%%%%%%
% Section 3.3
%%%%%%%%%%%%%%%%%%%%%%%
%
\subsection{Rotational variables from the Kepler field}
A large sample of rotational variables have recently been identified 
by McQuillan, Mazeh \& Aigrain~(\cite{mcquillan2014}) through analyzing 
the photometric database of the Kepler satellite. This includes all 
possible rotational variables in the Kepler field. In an earlier 
publication, McQuillan, Mazeh \& Aigrain~(\cite{mcquillan2013}) 
investigated only those targets that were labeled as Kepler objects 
of interest (KOI). Here we focus on this smaller sample, containing 
760 objects.\footnote{We attempted to also use the large sample of 
McQuillan, Mazeh \& Aigrain~(\cite{mcquillan2014}), but found that 
the errors in $\log g$ were so large (often an order of magnitude 
larger than for the field hot-Jupiter host stars) that a reasonable 
isochrone age determination was meaningless.} A negligible fraction 
($5$\%) of this sample contains probable eclipsing binaries, blends, 
or suffers from other ambiguities. 

Employing these KOI targets in our tests has the advantage of using 
photometric rotation periods instead of spectroscopic ones, thereby 
substantially decreasing the additional source of scatter due to the 
aspect angle dependence. However, there is also a drawback of using 
these data. Since the Kepler targets are numerous and have considerably 
lower apparent brightness, their basic spectroscopic and photometric 
parameters are usually less accurate than those of the individually 
studied bright planet host stars. 

Since the published data do not contain evolutionary (or any) age 
information, we computed isochrone ages according to the Yonsei-Yale 
models 
(Demarque et al.~\cite{demarque2004}).\footnote{http://www.astro.yale.edu/demarque/yyiso.html} 
With the aid of their interpolation routine and additional fine-grid 
interpolation, we established a dense isochrone grid for solar-scaled 
models without $\alpha$ element enhancement. The metallicity and age 
grids are uniform and cover the following values: 
$\{Z_i=0.001+(i-1)0.0005; i=1,2,...,80\}$ and 
$\{\log t_j=8.00+(j-1)0.02; \text{}  j=1,2,...,107\}$. Each of the 
downloaded isochrone contains $140$ $(\log T_{\rm eff},\log g)$ mesh 
points that we further increased by linear interpolation to $\sim 560$, 
leading to sufficiently dense sampling in all parameters to be matched 
to the spectroscopic data. The solar heavy element abundance for these 
models is equal to $0.018$. We minimized the following metric to select 
the best matching models 

%################
%  Eq. (7)
%################
%
\begin{equation}
D^2 = w(\Delta \log T_{\rm eff})^2 + (1-w)(\Delta \log g)^2 \hskip 2mm ,  
\end{equation}
where $\Delta$ stands for the difference between the grid point and 
the observed values. The weight $w$ was chosen to be $5/6$ and takes 
into consideration the smaller range of $log T_{\rm eff}$ values 
entering the matching procedure. For the observed values falling 
within the region spanned by the isochrones, we derived matching distances 
$D$ lower than $0.002$, usually close to $0.001$ or lower (indicating that 
we have a sufficiently densely interpolated set of models). We obtained 
a rough estimate on the error of the age by computing the standard 
deviation of all model values satisfying the $D<0.04$ criterion. 
This rather high cutoff for $D$ is set for the broad agreement of 
our error estimates with those found in the literature. In computing 
$\sigma({\rm age})$, the standard deviation of the ages satisfying 
the condition posed by $D$, we weighted the various age values by the 
inverse of the square of the corresponding distances 

%################
%  Eq. (8)
%################
%
\begin{equation}
\sigma^2({\rm age}) = \sum_{\rm k=1}^n w_k({\rm age}_k-\langle {\rm age}\rangle)^2 \hskip 2mm ,  
\end{equation}
where $w_k=(1/D_k^2)/\sum_{\rm j=1}^n 1/D_j^2$ and $n$ is the total 
number of grid points satisfying the $D<0.04$ criterion, and $D_j$ 
is the distance $D$ (see Eq.~(7)) of the j-th grid point from the target 
values. The average age $\langle {\rm age}\rangle$ was computed in a 
similar manner: $\langle {\rm age}\rangle = \sum_{\rm k=1}^n w_k{\rm age}_k$. 
(We note that in the tests we used the best-fitting model value as our age 
estimate rather than this average.) In a comparison with the ages found 
in the literature on the sample of hot-Jupiter host stars (Sect.~3.2), 
we found that except for a few outliers at young or old ages and for some 
$20$\% of stars with deviations $0.5$--$1.5$~Gyr, this simple fitting method 
yields ages that agree well with the published values within 
$\pm0.5$~Gyr. (See also Appendix B for an example of this compatibility 
on a small size of sample of hot Jupiters compiled by 
Maxted et al.~\cite{maxted2015}.)

%################
% Figure 9
%################
%
\begin{figure}
 \vskip -0mm
 \includegraphics[angle=-90,width=90mm]{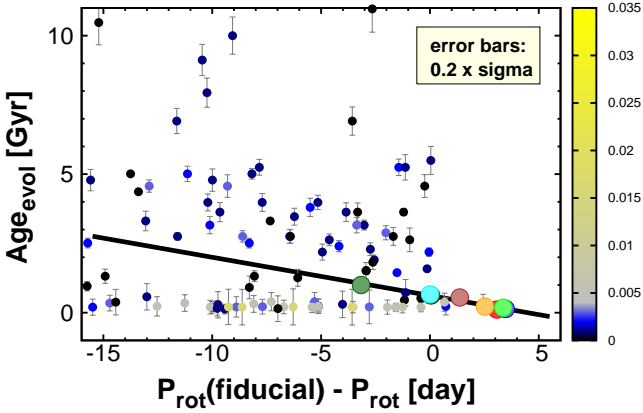}
 \vskip -0mm
 \caption{As in Fig.~\ref{evol-gyro-valenti}, but for the KOI sample  
          of McQuillan, Mazeh \& Aigrain~(\cite{mcquillan2013}). 
          The evolutionary ages were computed with the aid of 
          the Yonsei-Yale isochrones. The high-density horizontal 
          set of points at low isochrone ages dominantly constitutes 
          stars with observed stellar parameters outside the isochrone 
          regime. The color coding is scaled by the best isochrone 
          matching distance (Eq.~(7)) and shows that most of these 
          low-lying stars do not match the isochrones.}
\label{evol-gyro-koi}
\end{figure}

For the $762$ stars of McQuillan, Mazeh \& Aigrain~(\cite{mcquillan2013}) 
we used the spectroscopic and photometric parameters recently revised by 
Huber et al.~(\cite{huber2014}). To discard suspected items with excessive 
errors, we performed a fairly generous parameter cut by requiring both the 
[Fe/H] and the $\log g$ errors to be smaller than $0.2$. This filtering led 
to a sample of $207$ stars that was further decreased to $204$ by satisfying 
the condition of $D<0.04$. 

The resulting period$-$age plot is shown in Fig.~\ref{evol-gyro-koi}. 
The plot is very similar to the earlier ones, showing the significantly 
older ages obtained from the isochrone fits. In addition to this general 
feature, we also recognize a relatively large number of young (looking) 
objects with nearly the same age of $0.1$--$0.3$~Gyr. As the side bar 
shows, most of these objects have rather low isochrone matching accuracy 
(i.e., they have high $D$ values). Indeed, a comparison with the 
$\log T_{\rm eff} - \log g$ isochrone plot shows that nearly all of these 
objects are {\em outside} the regime covered by the isochrones for the 
given metallicity. More specifically, their spectroscopic gravity is 
higher than expected from the models. If we discard these objects, the 
trend toward greater isochrone ages is again clear.

%%%%%%%%%%%%%%%%%%%%%%%
% SECTION 4
%%%%%%%%%%%%%%%%%%%%%%%
%
\section{Conclusions}
The purpose of this paper was threefold: i) examine the validity of 
the Skumanich-type scaling between the stellar rotation period, color,
and age through a comprehensive analysis of the available observations 
on open clusters; ii) compare the ages based on the revised scaling 
with the stellar evolutionary (isochrone) ages for various samples 
of stars; and (iii) if the result of test (ii) is positive (i.e., no 
major discrepancy is found), recalibrate the gyro-age formula to 
accommodate the older ages and different evolutionary histories of 
these samples and thereby make the age determination based on stellar 
rotation more widely applicable.  

In test (i) we compiled the data from eight open clusters and searched 
for a simple transformation that generates a tight relation between 
the color and rotation period for the full sample. As has previously 
been indicated but left untreated in some other publications -- for 
example, Cargile et al.~\cite{cargile2014}, by using the Skumanich-type 
multiplicative period transformation, the periods are stretched too 
much at the long-period side (i.e., toward lower mass stars), leading 
to a rather fuzzy color$-$period plot when all data are combined. 
If we instead use a simple additive scaling (i.e., we shift the 
periods by cluster-dependent optimal constants), then the individual 
cluster ridges align in a much tighter way. This enabled us to 
investigate the relation between cluster ages and period shifts on a 
more solid statistical basis. The relation between these two quantities 
was also fairly tight, which together with the derived fiducial ridge 
(expressing the relation between the color and period for rotationally 
settled near main-sequence F--K dwarfs at a given age) enabled us to 
give age estimates for other stars based on their rotation periods and 
colors (see Eqs.~(3) and (5)).  

It is not the purpose of this paper to discuss the possible theoretical 
consequences of the better fitting additive period$-$age scaling. Here 
we only note three aspects that should be considered in constructing a 
revised model of stellar angular momentum dissipation. First, the 
rotational evolution of stars in open clusters might differ from those 
in the field due to the substantially different stellar environment and 
the low density of interstellar matter. Second, the rate of angular 
momentum loss is a strong function of the assumed structure of magnetic 
field and other physical details (e.g., core-envelope coupling, magnetic 
field strength vs. rotation, etc.) For example, 
Reiners \& Mohanty~(\cite{reiners2012}) derived a rotation-rate dependence 
of $t^{-1/4}$ in the non-saturated regime for $1M_{\odot}$ stars merely 
by adopting a different physical meaning of the magnetic flux -- rotation 
rate relation. Third, both the sample size and the age range of open 
clusters available currently for gyro-age studies are small, therefore 
the true age$-$rotational period relation might be more involved than 
the one derived here. (We sampled only a small part of an unknown function 
that was incidentally best approximated by a linear relation in this 
restricted parameter space.) 

Test (ii) resulted in a negative conclusion for all three datasets 
(field stars, host stars of transiting hot Jupiters, and Kepler planetary 
candidate stars). The bulk of the {\em isochrone ages \textup{is} 
$1.5$--$2.0$~{\rm Gyr} greater} than the predicted gyro-ages, with a 
large scatter to even larger differences. Only a small fraction of stars 
show younger isochrone ages in all three samples. It is important to 
recall that the ages of the open clusters -- which the gyro-ages have 
been calibrated to -- are also based {\em  on isochrone ages}, essentially 
on the {\em same} evolutionary stellar models, as those used in the 
test samples. Although the age overlap between the field stars and 
open cluster stars is not too extensive, it seems that the above difference 
is characteristic for all ages, also including the younger age range of 
the calibrating clusters. Therefore, the discrepancy does not seem to 
be the result of a poor extrapolation of the age relation derived from 
open clusters. Although the topology of the isochrones may introduce some 
bias toward older ages, this effect is likely to be small, based on the 
survival of the age discrepancy for stars with accurate isochrone ages 
(e.g., Sun, KELT-2A -- see Beatty et al.~\cite{beatty2012}). Apparently, 
non-cluster field stars have {\em significantly lower slow-down rates} 
than their cluster counterparts.  

This study supports the conclusions of other current works investigating 
the performance of the gyro-age method on various stellar populations. 
From the study of bright planet host stars, 
Maxted, Serenelli, \& Southworth~(\cite{maxted2015}) reached the same 
conclusion as we did, in spite of their quite different gyro-age method. 
Earlier, Brown~(\cite{brown2014}) also found hints of the young gyro-ages 
of extrasolar planet host stars. Based on a Monte Carlo study of a large 
sample of Kepler asteroseismic targets and data from two open clusters, 
Angus et al.~(\cite{angus2015}) also questioned the overall reliability 
of the current gyro-age estimations. 

In these circumstances, we were obviously unable to pursue task (iii). 
A more reliable extension of the gyro-age method to non-cluster stars 
should probably wait until the source of the discrepancy between the 
current gyro- and isochrone ages of the field and cluster stars is 
understood and a physically acceptable solution is found. Finally, we 
note that our conclusion is based on isochrone ages derived from 
non-rotating stellar evolutionary models. Brandt \& Huang~(\cite{brandt2015}) 
recently showed that with rotating models the age of the Hyades and 
Praesepe clusters become considerably older (from the generally accepted 
value of 630/670~Myr -- which we also used here -- to $\sim 800$~Myr). 
An extension of the evolutionary models in this direction might mitigate 
some part of the discrepancy we highlighted here.

\begin{acknowledgements}
This work has been started during G.~K.'s stay at the Physics and 
Astrophysics Department of the University of North Dakota. He is 
indebted to faculty and staff for the hospitality and the cordial, 
inspiring atmosphere. I am greatly indebted to Joel Hartman for 
sharing his manifold expertise on the subject and for his valuable 
suggestions and comments during the preparation of the paper. 
Critical (but constructive) comments by the anonymous referee were 
instrumental in the final shaping of the paper. This publication 
makes use of the SIMBAD database and the VizieR catalogue access 
tool, operated at CDS, Strasbourg, France. This research was made 
possible through the use of the AAVSO Photometric All-Sky Survey 
(APASS), funded by the Robert Martin Ayers Sciences Fund.  
\end{acknowledgements}

\bibliographystyle{aa} % style aa.bst

%
%%%%%%%%%%%%%%%
%  APPENDIX A 
%%%%%%%%%%%%%%%
%

\begin{appendix}
\section{Fiducial polynomial and the robust fit of the `I' sequences}
\begin{figure}
\center
 \vskip -0mm
 \includegraphics[angle=-90,width=70mm]{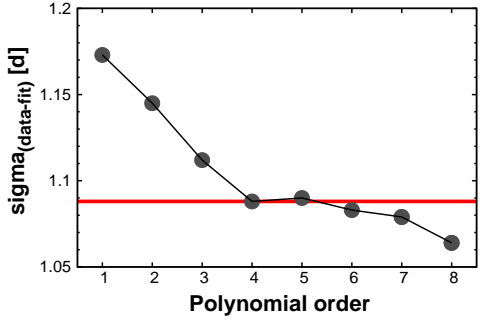}
 \vskip -0mm
 \caption{Polynomial order vs. unbiased estimate of the standard 
       deviation of the residuals of the fit for the fiducial 
       ridge fitting (see Fig.~\ref{polynomial_fit}). For reference, 
       the horizontal line shows to the standard deviation at order 
       four.}
\label{order-sigma}
\end{figure}

To find the best representation of the joint `I' sequences of the merged 
data of Praesepe (M44) and the Hyades, we fitted polynomials of various 
order and checked the fit both by visual and by statistical means. We 
omitted the obvious outliers that are mostly due to the rotationally 
unsettled lower temperature stars in the Hyades (we note that all 
outliers are well defined). Altogether, we left out $\text{two}$ stars 
from Praesepe and $11$ from the Hyades and compiled a sample of $228$ 
stars. These stars were fitted by least squares of equal weights. 
Figure~\ref{order-sigma} shows the variation of the unbiased estimate 
of the standard deviation of the residuals as a function of the polynomial 
order. 

The standard deviation levels off at about order four. Although there 
is some decrease afterward, the fit shows wiggles with increasing 
amplitudes as the order of the polynomial increases. The fit starts to 
become unstable at order nine, and it becomes entirely volatile at order 
ten with a residual standard deviation of $4.1$. We also tested the 
statistical significance of the fourth-order fit by using the following 
statistics

%################
%  Eq. (A.1)
%################
%
\begin{eqnarray}
R(j_1,j_2) & = & \left({{RSS(j_1)} \over {RSS(j_2)}} - 1\right){{n-j_2} \over {j_2-j_1}} \nonumber\\
RSS_k & = & \sum_{i=1}^n (P_{rot}(i)-p_{k}(i))^2 \hskip 2mm .
\end{eqnarray}
Here $n$ is the number of data points, $\{p_{k}(i)\}$ is the fitted 
polynomial of order $k$, and $j_1$ and $j_2$ are the order tested. In our 
case, $j_2=j_1+1$, so $R(j_1,j_2)$ follows a Fisher distribution of 
$F(1,n-j_2)$. We assessed the significance of the change in 
$R(j_1,j_2)$ in terms of the theoretical standard deviation of 
$F(1,n-j_2)$. This yields some $8\sigma$ significance at each step 
for the change in $R(j_1,j_2)$ as we go from the second- to the 
fourth-order fits. After this, the significance decreases to 
$\sim 2\sigma$ with an increase at order eight and then the solution 
becomes unstable, as mentioned.  

\begin{figure}
\center
 \vskip -0mm
 \includegraphics[angle=-90,width=70mm]{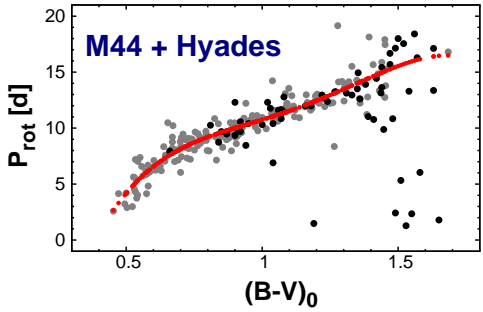}
 \vskip -0mm
 \caption{Fourth-order polynomial fit to the combined data of 
        M44 (Praesepe, gray points) and the Hyades (black points). 
        The fit (red dots) is sampled at the same $(B-V)_0$ values 
        as the data. All data are plotted, including the $13$ 
        outliers mentioned in the text.}
\label{polynomial_fit}
\end{figure}

The fourth-order polynomial fitted to the merged data of M44 and the 
Hyades is shown in Fig.~\ref{polynomial_fit}. The regression parameters 
and their statistical errors are given by Eq.~\ref{fiducial_pol}. 

\begin{figure}
\center
 \vskip -0mm
 \includegraphics[angle=-90,width=70mm]{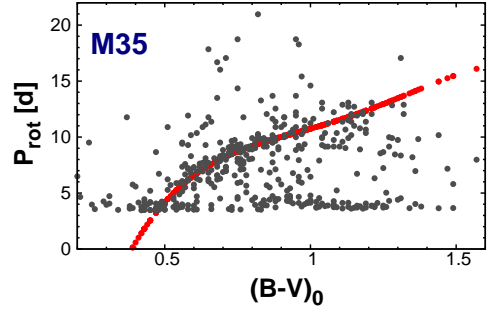}
 \vskip -0mm
 \caption{Result of the best-fit period shift of the rotation periods 
       of M35 (gray dots) to the fiducial ridge (red dots) by using 
       the GM kernel (German \& McClure~\cite{german1987}). The 
       fiducial ridge is sampled at the same $(B-V)_0$ values as the 
       data.}
\label{robust_fit}
\end{figure}

This fiducial polynomial was used in Sect.~2.2 to derive the 
relative period shifts of the `I' sequences of the individual 
clusters. The best-fit shifts were determined by using a 
kernel-weighted least-squares method to consider outliers (i.e., 
minimizing the effect of stars not associated with the `I' 
sequence). The kernel proposed by German \& McClure~(\cite{german1987})  
has proven to be an excellent way of localizing the ridge in 
each cluster. We show an example for the robustness of the fit in 
Fig.~\ref{robust_fit}. 

%################
% Figure 10
%################
%

%################
% Figure 11
%################
%

%################
% Figure 12
%################
%
\end{appendix}

%
%%%%%%%%%%%%%%%
%  APPENDIX B 
%%%%%%%%%%%%%%%
%

\begin{appendix}
\section{Comparison with other studies}
\begin{figure}
\center
 \vskip -0mm
 \includegraphics[angle=-90,width=90mm]{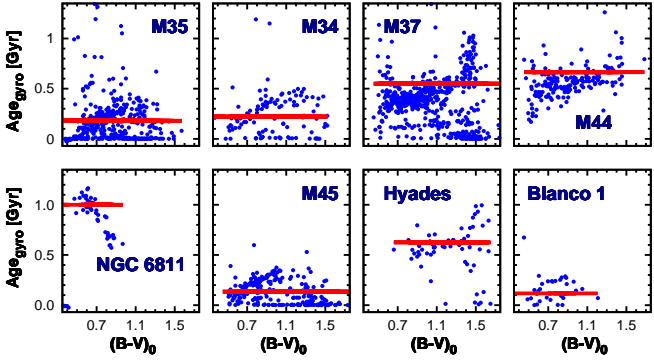}
 \vskip -10mm
 \caption{Predicted individual gyro-ages (dots) by using the formula of 
       Angus et al. (2015) and the adopted isochrone cluster ages of 
       Table~\ref{clusters} (lines).}
\label{mult_ages}
\end{figure}

First we calculated the individual rotational ages of the cluster sample 
used in this paper by employing the gyro-age formula of 
Angus et al.~(\cite{angus2015}). Their Eq.~(15) has the standard form as 
introduced by Barnes~(\cite{barnes2003}), but the parameters are determined 
by using $310$ astroseismic targets from the survey of the Kepler satellite 
and some additional stars from the field and from some open clusters. We 
plot the individual gyro-ages for the eight clusters in Fig.~\ref{mult_ages}. 
In comparison with Fig.~\ref{gyro-iso-clusters}, using our additive 
age-period-color relation, the ages corresponding to the stars associated 
with the ridges of type `I' have stronger systematic variations than those 
derived from our additive formula (see, e.g., the plot for M37). This is 
the effect of multiplicative period--age scaling, as discussed in Sect.~2.   

In the rest of this appendix we show that the significantly younger 
rotational ages derived for the field stars from our newly calibrated 
additive formula is not specific to this formula, but is a general property 
of all currently used gyro-age calibrations. 

\begin{figure}
\center
 \vskip -0mm
 \includegraphics[angle=-90,width=55mm]{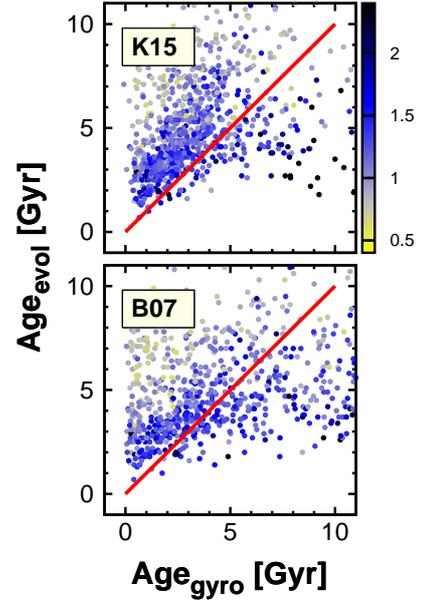}
 \vskip -0mm
 \caption{Rotational ages vs. stellar evolution ages computed using the additive 
          gyro-age formula Eq.~(5) derived in this paper (upper panel) and 
          the popular multiplicative formula of Barnes~(\cite{barnes2007}) 
          (lower panel). The thick continuous lines indicate the identical 
          values of the rotational and evolutionary ages. We use the 
          spectroscopic data of Valenti \& Fisher~(\cite{valenti2005}). The 
          color coding is for the stellar radius as indicated by the side bar 
          (scaled in solar units). 
          }
\label{k15-b07}
\end{figure}

In the lower panel of Fig.~\ref{k15-b07} we plot the evolutionary ages as 
given in the spectroscopic survey of $1039$ field stars by 
Valenti \& Fisher~(\cite{valenti2005}) vs. the rotational ages derived from 
one of the most frequently used gyro-age formula of Barnes~(\cite{barnes2007}, 
his Eq.~(1)). For comparison, in the upper panel of the same figure we show 
the plot derived from the additive gyro-age formula presented in this paper. 
In both cases we use a subsample of the full survey as described in Sect.~3.1. 

Although different in details, the two plots are topologically closely 
similar, indicating in both cases a significant excess of old evolutionary 
ages. The high density of points in the [0,5]~Gyr regime suggests an overall 
difference of 1--2~Gyr. From the color-coded radius distribution it is also 
clear that differences much higher than the quoted value exist for stars 
with lower radii. By repeating the iso-gyro age comparison for other empirical 
gyro-age formulae -- that is, those of Mamajek \& Hillenbrand~(\cite{mamajek2008}) 
(their Eqs. (12)--(14) with the parameters given in their Table (10)) and 
Angus et al.~(\cite{angus2015}) (their Eq.~(15)) -- a similar conclusion can be 
drawn on the topology of the gyro- vs. evolutionary age relation. Interestingly, 
these two works yield very similar parameters for the Barnes-type gyro-age 
relation, in spite of the very different input data and methods of analysis, 
and the conclusion of the authors in the second paper about the limited 
applicability of the formulae presented for the full population of the dataset 
used in the calibration.    

\begin{figure}
\center
 \vskip -0mm
 \includegraphics[angle=-90,width=55mm]{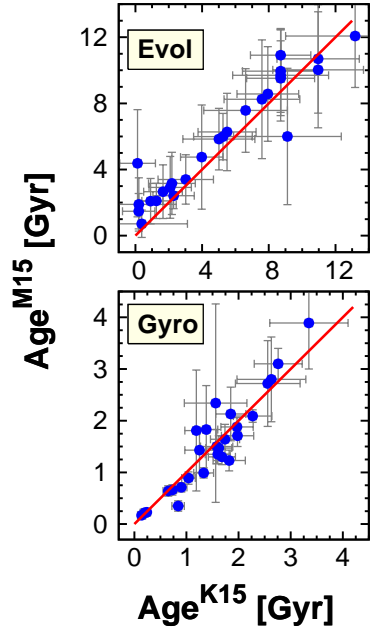}
 \vskip -0mm
 \caption{Comparison of the evolutionary ages (upper panel) and the 
 rotational ages (lower panel) for the extrasolar planet host 
 stars of Maxted~et. al~(\cite{maxted2015}) with those derived in 
 this paper using the Yonsei-Yale isochrones of 
 Demarque et al.~(\cite{demarque2004}) and the additive gyro-age formula 
 Eq.~(5) of this paper. The continuous line indicates the identical 
 values for the quantities corresponding to the labels. We note that the panels have
 different ranges.}
\label{k15-m15}
\end{figure}

Yet another approach for estimating gyro-ages comes from the analytical model 
of Barnes~(\cite{barnes2010}, see also Barnes \& Kim~\cite{barnes-kim2010}). 
Maxted et al.~(\cite{maxted2015}) employed this model to analyze $28$ 
well-established extrasolar planet host stars with measured photometric 
rotation periods. We employed this sample with the stellar parameters used 
in their paper to compute our age estimates. The plots in Fig.~\ref{k15-m15} 
show the excellent correlations between the ages of 
Maxted et al.~(\cite{maxted2015}) and those presented here.\footnote{We note, 
however, that 55~Cnc is not plotted in this graph. The gyro-ages derived for 
this object differ considerably ($8.10\pm3.54$ for 
Maxted et al.~\cite{maxted2015} and $4.62\pm1.49$ for our formula). Although 
the difference is within the error limit, the discrepancy between the two 
types of gyro-ages indicates that they show very different behavior for long 
rotational periods. For example, HATS-2 has very similar stellar parameters 
to those of 55~Cnc, but a far shorter rotational period 
($24.98\pm0.04$ vs. $39.0\pm9.0$ days). Their gyro-ages agree fairly closely 
($3.10\pm0.30$ and $2.76\pm0.46$ for Maxted et al.~\cite{maxted2015} and 
this paper, respectively). The discrepancy for 55~Cnc might disappear in 
the future, once the estimate of its rotation period becomes more accurate 
with a presumed shift toward shorter periods.} 
This good agreement is rather surprising for the gyro-ages, since our ages 
are basically empirical (with the intermediation of the cluster ages determined 
by the stellar evolutionary isochrone fits), whereas their ages are more involved 
through various model approximations and initial conditions. For the evolutionary 
ages there is a uniform shift of $0.5$--$1$~Gyr in the sense that the ages 
computed by Maxted et al.~(\cite{maxted2015}) are older than ours. This tendency 
of the {\sc garstec} models of Weiss \& Schlattl~(\cite{weiss2008}) used by 
Maxted et al.~(\cite{maxted2015}) was also noted by Metcalfe et al.~(\cite{metcalfe2014}). 
Chaplin et al.~(\cite{chaplin2014}) attributed this offset to the different 
treatment of convective core overshooting in the {\sc garstec} models. 

%################
% Figure 13
%################
%

%################
% Figure 14
%################
%

%################
% Figure 15
%################
%

\end{appendix}

\end{document}